\documentclass[letterpaper,showpacs,superscriptaddress,amsfonts,amssymb,reqno]{amsart}

\numberwithin{equation}{section}

\usepackage{amsmath}
\usepackage{color}

\newcommand{\mc}[1]{{\mathcal #1}}

\newcommand{\mb}[1]{{\mathbb #1}}

\renewcommand{\epsilon}{\varepsilon}

\title{Quantitative analysis of Clausius inequality}

\author [L. Bertini] {Lorenzo Bertini}
\address{\noindent Lorenzo Bertini \hfill\break\indent 
Dipartimento di Matematica, Universit\`a di Roma `La Sapienza' 
\hfill\break\indent 
P.le Aldo Moro 2, 00185 Roma, Italy}
\email{bertini@mat.uniroma1.it}

\author [A. De Sole] {Alberto De Sole}
\address{\noindent Alberto De Sole \hfill\break\indent 
Dipartimento di Matematica, Universit\`a di Roma `La Sapienza' 
\hfill\break\indent 
P.le Aldo Moro 2, 00185 Roma, Italy}
\email{desole@mat.uniroma1.it}

\author[D. Gabrielli]{Davide Gabrielli}
\address{\noindent Davide Gabrielli \hfill\break\indent 
 Dipartimento di Matematica, Universit\`a dell'Aquila
\hfill\break\indent 
67100 Coppito, L'Aquila, Italy
}
\email{gabriell@univaq.it}

\author[G. Jona-Lasinio]{Giovanni Jona-Lasinio}
\address{\noindent Giovanni Jona-Lasinio
\hfill\break\indent 
 Dipartimento di Fisica and INFN, Universit\`a di Roma La Sapienza
\hfill\break\indent 
 P.le A.\ Moro 2, 00185 Roma, Italy
}
\email{gianni.jona@roma1.infn.it}

\author[C. Landim]{Claudio Landim} 
\address{Claudio Landim
  \hfill\break\indent IMPA \hfill\break\indent Estrada Dona Castorina
  110, \hfill\break\indent
J. Botanico, 22460 Rio de Janeiro, Brazil\hfill\break\indent
  {\normalfont and} \hfill\break\indent CNRS UMR 6085, Universit\'e de
  Rouen, \hfill\break\indent Avenue de l'Universit\'e, BP.12,
  Technop\^ole du Madril\-let, \hfill\break\indent
F76801 Saint-\'Etienne-du-Rouvray, France.} 
\email{landim@impa.br}

\begin{document}

\begin{abstract}
  In the context of driven diffusive systems, for thermodynamic
  transformations over a large but finite time window, we derive an 
  expansion of the energy balance.
  In particular, we characterize the transformations which minimize the
  energy dissipation and describe the optimal correction to the quasi-static
  limit. Surprisingly, in the case of transformations between homogeneous
  equilibrium states of an ideal gas,
  the optimal transformation is a sequence of inhomogeneous
  equilibrium states.
\end{abstract}


\maketitle
\thispagestyle{empty}

\section{Introduction}

As discussed in thermodynamic textbooks, in a transformation
between equilibrium states a system necessarily goes through
deviations from equilibrium which are small if the transformation is
quasi-static. As clearly stated in Callen \cite{callen},
\begin{quote}
    A quasi-static process is thus defined in terms of a dense
    succession of equilibrium states. It is to be stressed that a
    quasi-static process therefore is an idealized concept, quite
    distinct from a real physical process, for a real process always
    involves nonequilibrium intermediate states having no
    representation in the thermodynamic configuration
    space. Furthermore, a quasi-static process, in contrast to a real
    process, does not involve considerations of rates, velocities or
    time. The quasi-static process simply is an ordered succession of
    equilibrium states, whereas a real process is a temporal
    succession of equilibrium and nonequilibrium states.
\end{quote}
Our aim is to develop the analysis, started in \cite{45,46}, of real
transformations for driven diffusive systems, both in equilibrium and
nonequilibrium. As emphasized in the previous quotation this
analysis will necessarily involve dynamical considerations that are
outside the scope of classical thermodynamics.

The dynamic evolution of driven diffusive system is described by the
continuity equation together with the constitutive equation that
expresses the local current as a function of the density and the
driving field. The interaction with boundary reservoirs specifies the
appropriate boundary conditions. A (real) transformation is thus defined
by a choice of time-dependent driving field and chemical potentials of
the reservoirs. Within this scheme, a dynamical derivation of the
Clausius inequality $W\ge \Delta F$ for isothermal transformations has
been obtained in \cite{45,46}. Here $W$ is the work done in the
transformation and $\Delta F$ is the variation of the free energy.
Moreover, we have shown that the Clausius inequality becomes an equality,
i.e., $W=\Delta F$ in the limit of very slow transformations, that is
in the quasi-static limit.

Since real transformations last a finite time, the quasi-static limit
cannot be achieved. A meaningful issue is thus to describe the
corrections to the quasi-static limit for transformations over a large
but finite time window $\tau$. To describe the evolution of the system
it is convenient to rescale time by introducing the dimensionless
variable $s=t/\tau$ where $t$ is the original time variable. We then
expand the evolution and the energy balance in powers of $1/\tau$ and
compute the first order corrections.

Consider now transformations through equilibrium states
namely, for which the stationary current, corresponding to the given external drivings at time $s$, 
vanishes.
In absence of an external field,
transformations through equilibrium states are those in which 
the chemical potentials, while varying in time. are the same on each point of the boundary. 
For transformations through equilibrium states we show that, up
to order $1/\tau^2$, $W= \Delta F + (1/\tau) \,  B$ where $B$ is a positive
functional of the transformation,
that is of the time dependent external drivings.
For real but slow transformations we
can thus optimize the dissipated energy  by minimizing the functional $B$. 
Not surprisingly,  we prove that for transformations between equilibrium states 
(namely such that the initial and final states are equilibrium states), the
functional $B$ is minimized by transformations through equilibrium
states.

In the case of an ideal gas, the minimizer of $B$ can be
computed explicitly. Somehow surprisingly, for transformation between
homogeneous equilibrium states (characterized by the absence of
external field), the optimal transformation is a
sequence of inhomogeneous equilibrium states. In other words, it is
profitable to switch on an external field.  
In the context of Langevin dynamics, finite time refinements to the
second law of thermodynamics have been discussed in \cite{AGMMM}, see
also \cite{GMP} for the case of jump Markov processes.

For transformations between nonequilibrium states, the Clausius
inequality $W\ge \Delta F$ does not carry any significant information.
In fact, the energy dissipated along
such transformations will necessarily include the contribution needed
to maintain the nonequilibrium stationary states, which is infinite in
an unbounded time window.
It is however possible to formulate a meaningful version of the Clausius
inequality for nonequilibrium states by introducing a
\emph{renormalized work} $W^\mathrm{ren}$ that is defined by subtracting from the total
work $W$ the energy needed to maintain the nonequilibrium state.  Within
the setting of the macroscopic fluctuation theory \cite{rmp}, a
definition of renormalized work has been proposed in \cite{45,46}
following the point of view in \cite{op} further developed in
\cite{hs,kn,knst}.
The analysis of real thermodynamic transformations carried out in this
paper and outlined above includes transformations between
nonequilibrium states provided the work is replaced by the
renormalized work.

We draw the reader's attention to the very recent paper \cite{mj}.
This paper analyses,
in the context of Markov chains with finitely many degrees of freedom,
problems similar to the ones discussed here.
In particular, the authors introduce an optimization problem for the finite time correction
to Clausius inequality, 
with motivations similar to ours.

\section{Clausius inequality and its nonequilibrium counterparts}

In this section we review the dynamical approach to thermodynamic transformations 
introduced in \cite{45,46} and developed in \cite{rmp}.

\subsection*{Hydrodynamical description}
We introduce the hydrodynamic description of out of equilibrium
driven diffusive systems which are characterized by conservation
laws. 
We restrict to the case of a single conservation law, e.g., the
conservation of the mass. 

 We denote by $\Lambda \subset \mb R^d$ the
bounded region occupied by the system, by $\partial \Lambda$ the
boundary of $\Lambda$, by $x$ the macroscopic space coordinates and by
$t$ the macroscopic time. The system is in contact with boundary
reservoirs, characterized by their chemical potential $\lambda (t,x)$,
and under the action of an external field $E(t,x)$.

At the macroscopic level the system is completely described
by the local density $\rho(t,x)$ and the local density current $j(t,x)$. Their
evolution is given by the continuity equation together with the constitutive
equation which expresses the current as a function of the
density. Namely,
\begin{equation}
\label{2.1}
\begin{cases}
\partial_t \rho (t) + \nabla\cdot j (t) = 0,\\
j (t)= J(t,\rho(t)),
\end{cases}
\end{equation}
where we omit the explicit dependence on the space variable $x\in\Lambda$.
For driven diffusive systems the constitutive equation takes the form
\begin{equation}
\label{2.2}
J(t,\rho)  = - D(\rho) \nabla\rho + \chi(\rho) \, E(t),
\end{equation}
where the \emph{diffusion coefficient} $D(\rho)$ and the \emph{mobility}
$\chi(\rho)$ are assumed to be $d\times d$ symmetric and positive definite matrices.
This holds in the context of stochastic lattice gases \cite{splib}. 
Equation \eqref{2.2} relies on the diffusive approximation and on the
linear response to the external field.
The evolution of the density is thus given  by the driven diffusive
equation
\begin{equation}
\label{r01}
\partial_t \rho (t) + \nabla\cdot \big( \chi(\rho)  E(t) \big)
= \nabla\cdot \big( D(\rho) \nabla\rho \big).
\end{equation}
 
The transport coefficients $D$ and $\chi$ satisfy the local Einstein relation
\begin{equation}
\label{r29}
D(\rho) = \chi(\rho) \, f''(\rho),
\end{equation}
where $f$ is the equilibrium free energy per unit  volume.

Equations \eqref{2.1}--\eqref{2.2} have to be supplemented by the
appropriate boundary condition on $\partial\Lambda$ due to the
interaction with the external reservoirs. If $\lambda(t,x)$,
$x\in\partial \Lambda$, is the chemical potential of the external
reservoirs, the boundary condition reads
\begin{equation}
\label{2.3}
f'\big(\rho(t,x) \big) = \lambda(t,x), \qquad x\in\partial \Lambda.
\end{equation}

\medskip
If the chemical potential and external field do not depend on time, we
denote by $\bar\rho=\bar\rho_{\lambda, E}$ the stationary solution of
\eqref{r01},\eqref{2.3},
\begin{equation}
\label{05}
\begin{cases}
\nabla \cdot J(\bar\rho)= \nabla \cdot \big( -D(\bar\rho)
\nabla\bar\rho + \chi(\bar\rho) \, E  \big) = 0,  
\\
 f' (\bar\rho(x)) = \lambda (x), 
\qquad x\in\partial \Lambda.
\end{cases}
\end{equation}
We will assume that this stationary solution is unique.
The stationary density profile $\bar\rho$ is characterized by the
vanishing of the divergence of the associated current, $\nabla\cdot
J(\bar\rho)=0$.  A special situation is when the current itself
vanishes, $J(\bar\rho)=0$; if this is the case we say that the system
is in an equilibrium state; this can be viewed as a macroscopic counterpart to 
detailed balance.
Conversely given a density profile $\bar\rho$ there is not a unique pair $(\lambda,E)$
such that $\bar\rho=\bar\rho_{\lambda,E}$. Indeed $\lambda$ is uniquely determined by
the second equation in \eqref{05} while the external field can be chosen in the form
\begin{equation}
E=\chi^{-1}(\bar\rho)\big(G+D(\bar\rho)\nabla\bar\rho\big)\,,
\label{per-alberto}
\end{equation}
where $G$ is an arbitrary divergence free vector field.
We note that for equilibrium states there is a one-to-one correspondence
between the pair $(\lambda, E)$ and the stationary solution of
\eqref{05}, that is defined by choosing $G=0$ in \eqref{per-alberto}.

Homogeneous equilibrium states correspond to the case in which the
external field vanishes and the chemical potential is constant. The
stationary solution is then constant and satisfies
$f'(\bar\rho_{\lambda,0}) =\lambda$.  Inhomogeneous equilibrium states
correspond to the case in which the external field is gradient,
$E=-\nabla U$, and it is possible to choose the arbitrary constant in
the definition of $U$ such that $U(x)=-\lambda(x)$,
$x\in\partial\Lambda$. 
By the Einstein relation \eqref{r29}, the stationary solution satisfies
$-f'\big(\bar\rho_{\lambda,E}(x)\big)=U(x)$ and the stationary current
vanishes, $J(\bar\rho_{\lambda,E})=0$.  Examples of inhomogeneous
equilibrium states in presence of an external field are provided by a
still atmosphere in the gravitational field or by sedimentation in a
centrifuge.

\medskip
In this framework it is possible to define a thermodynamic functional $V$,
called the quasi-potential,
generalizing the free energy for systems out of equilibrium.
It can be characterized 
as the maximal positive solution, vanishing when $\rho=\bar \rho$,
of the Hamilton-Jacobi equation,
\begin{equation}
\label{r15-2}
\int_{\Lambda}dx\,  \nabla \frac{\delta V}{\delta \rho} \cdot \chi(\rho)
\nabla \frac{\delta V}{\delta \rho}
-  \int_{\Lambda}dx\, \frac{\delta V}{\delta \rho} \nabla \cdot J(\rho)
= 0\,.
\end{equation}

We now define the \emph{symmetric current} $J_S$ by 
\begin{equation}
  \label{Ons-form}
  J_S(\rho)=-\chi(\rho)\nabla\frac{\delta V}{\delta\rho}.
\end{equation}
Since the stationary density $\bar\rho$ is a minimum for $V$, then
$(\delta V/\delta\rho)(\bar\rho)=0$.  The symmetric current thus
vanishes at the stationary profile,
\begin{equation}
\label{01}
J_S(\bar\rho) \;=\;0.
\end{equation}
We rewrite the hydrodynamic current as
\begin{equation}
\label{split}
J(\rho)=J_S(\rho)+J_A(\rho), 
\end{equation}
which defines the \emph{antisymmetric current} $J_A$. 

In view of these definitions, equation \eqref{r15-2} becomes
\begin{equation}
\label{r18}
\int_{\Lambda}dx\,  J_S (\rho) \cdot \chi(\rho)^{-1} J_A(\rho)
= 0.
\end{equation}

In the case of an equilibrium state the quasi-potential
$V=V_{\lambda,E}(\rho)$ is the local functional
\begin{equation}\label{localV}
V_{\lambda,E}(\rho)
=
\int_\Lambda dx
\big(
f(\rho)-f(\bar\rho)-f^\prime(\bar\rho)(\rho-\bar\rho)
\big)
\,,
\end{equation}
where $\bar\rho=\bar\rho_{\lambda,E}$ is the  solution of \eqref{05}.

\subsection*{Clausius inequality}
\label{s:rci}
The second law of thermodynamics can be expressed as follows. Consider
a system in an equilibrium state in thermal contact with an
environment at a given tempera\-ture. The system then undergoes an
isothermal transformation to a final state. 
The Clausius inequality states
\begin{equation}
  \label{ci}
  W \ge \Delta F ,
\end{equation}
where $W$ is the mechanical work done on the system
and $\Delta F$ is the difference of the free energy between
the final and the initial state. If equality holds the transformation is said
to be reversible. It can be implemented by performing very slow
variations so that the system goes through a sequence of equilibrium
states.  

We review the dynamical derivation of the Clausius inequality in
\cite{45,46}.
Consider a system in a time dependent environment, that
is, $E$ and $\lambda$ depend on time. The work done by the environment on the system in the
time interval $[0,T]$ is
\begin{equation}
  \label{W=}
  \begin{split}
    W_{[0,T]} = & \int_{0}^{T}\! dt \, \Big\{ \int_\Lambda \!dx\, j(t)
    \cdot E(t)
    - \int_{\partial\Lambda} \!d\sigma \, \lambda
    (t) \: j(t) \cdot \hat{n} \Big\},
  \end{split}
\end{equation}
where $\hat n$ is the outer normal to $\partial \Lambda$ and $d\sigma$
is the surface measure on $\partial \Lambda$.  The first term on the
right hand side is the energy provided by the external field
while the second is the energy provided by the reservoirs.

Fix time dependent paths $\lambda(t)$ of the chemical potential and
$E(t)$ of the driving field. Given a density profile $\rho_0$, let
$\rho(t)$, $j(t)$, $t \ge 0$, be the solution of
\eqref{2.1}--\eqref{2.3} with initial condition $\rho_0$.  
By using the Einstein relation \eqref{r29} and the boundary condition 
$f'(\rho(t)) = \lambda(t)$, an application of the
divergence theorem yields
\begin{equation}
\label{04}
W_{[0,T]} \,=\,  F(\rho(T)) - F(\rho(0)) 
+\,  \int_{0}^{T} \!dt  \int_\Lambda \!dx \;
j(t)\cdot \chi(\rho(t) )^{-1} j(t),
\end{equation}
where $F$ is the equilibrium free energy functional,
\begin{equation}
\label{10}
F(\rho) = \int_\Lambda \!dx \: f (\rho(x)).
\end{equation}
Equation \eqref{04} is not simply a rewriting of
\eqref{W=}, as it depends on a physical principle, the local
Einstein relationship.

Since the second term on the right hand side of \eqref{04} is
positive, we deduce the Clausius inequality \eqref{ci} with $\Delta F
= F(\rho_1) - F(\rho_0)$ for arbitrary density profiles
$\rho_0=\rho(0)$, $\rho_1=\rho(T)$.  Note that this derivation holds
both for equilibrium and nonequilibrium systems.

\subsection{Renormalized work}
The idea to define a renormalized work is to subtract 
the energy needed to maintain the system out of equilibrium.
For time independent drivings, by 
the orthogonal decomposition \eqref{split} and \eqref{01},
$J(\bar\rho)=J_{\mathrm{A}} (\bar\rho)$ is the macroscopic current in
the stationary state.  In view of the general formula for the total
work \eqref{04}, the amount of energy per unit time needed to maintain
the system in the stationary profile $\bar\rho$ is
\begin{equation}
\label{toman}
\int_\Lambda \!dx \:
J_\mathrm{A}(\bar\rho) \cdot \chi(\bar\rho)^{-1}
J_\mathrm{A}(\bar\rho).
\end{equation}

Fix now $T>0$, a density profile $\rho_0$, and space-time dependent
chemical potentials $\lambda(t)$ and external field $E(t)$, $t\in
[0,T]$.  Let $(\rho(t),j(t))$ be the corresponding solution of
\eqref{2.1}--\eqref{2.3} with initial condition $\rho_0$.  
We define the renormalized work
$W^\textrm{ren}_{[0,T]}$ done by the reservoirs and the external field
in the time interval $[0,T]$ as
\begin{equation}
\label{Weff}
W^\textrm{ren}_{[0,T]} = W_{[0,T]}
- \int_{0}^{T}\! dt \int_\Lambda \!dx\, J_\mathrm{A}(t,\rho(t)) \,
\cdot \chi(\rho(t))^{-1} J_\mathrm{A}(t,\rho(t))
\end{equation}
where $J_\mathrm{A}(t,\rho)$ is the antisymmetric current for the
system with the time independent external driving obtained by freezing
the time dependent chemical potential $\lambda$ and external field $E$
at time $t$.  Observe that the definition of the renormalized
work involves the antisymmetric current $J_\mathrm{A}(t)$ computed not
at density profile $\bar\rho_{\lambda(t), E(t)}$ but at the solution
$\rho(t)$ of the time dependent hydrodynamic equation.

The definition \eqref{Weff} is natural within the macroscopic
fluctuation theory and leads to a Clausius inequality.  Indeed, in
view of \eqref{04} and the orthogonality in \eqref{r18} between the
symmetric and the antisymmetric part of the current,
\begin{equation}
\label{Weff2}
  \begin{split}
    W^\textrm{ren}_{[0,T]} & = F(\rho(T)) - F(\rho_0)
    + \int_{0}^{T}\! dt \int_\Lambda \!dx\, J_\mathrm{S}(t,\rho(t)) \cdot
    \chi(\rho(t))^{-1} J_\mathrm{S}(t,\rho(t))\\
    & \geq F(\rho(T)) - F(\rho_0)\,.
  \end{split}
\end{equation}

In the context of Langevin dynamics, a different definition of
renormalized work has been proposed in \cite{mn},
see \cite{46} for a comparison.

We obtain next a macroscopic version of the well known Hatano-Sasa inequality \cite{hs}.
Consider the dissipation due to the symmetric current
\begin{equation*}
\int_{0}^{T} \!dt  \int_\Lambda \!dx \;
J_S(t,\rho(t))\cdot \chi(\rho(t) )^{-1} J_S(t,\rho(t))=
-\int_{0}^{T} \!dt  \int_\Lambda \!dx \;\nabla\frac{\delta V_t}{\delta\rho}\cdot J_S(t,\rho(t)) \geq 0 
\,,
\end{equation*}
where $V_t$ is the quasi-potential corresponding to the driving $\alpha=(\lambda(t),E(t))$
at frozen time $t$.
Integrating by parts and using the orthogonality between $J_S$ and $J_A$
we obtain
\begin{equation}
\int_0^Tdt \int_\Lambda \; dx \;\dot \alpha \cdot{\frac {\delta V_t}{\delta \alpha}} \geq V_T(\rho(T))-V_0(\rho(0))
\end{equation}
If the initial state is the stationary profile for $(\lambda(0),E(0))$, then the right hand side is $V_T(\rho(T)) \geq 0$.

\section{Finite time thermodynamics}
\label{FTT}

In this section we develop an approach to thermodynamic
transformations which takes into account the fact that any real
transformation lasts for a finite time.  We consider transformations
over an interval of time $[0,\tau]$ and we discuss their asymptotic
properties for large $\tau$.  In particular, for slow transformation,
we shall obtain the correction of order $\frac1\tau$ to the equality
$W=\Delta F$ which holds in the quasi-static limit. We finally discuss
which transformations minimize such a correction.

\subsection*{Slow transformations}

To analyze transformations over the interval $[0,\tau]$ it is
convenient to introduce the dimensionless variable $s={t}/{\tau}$.
A \emph{protocol} is defined by a choice of the external drivings
$E(s,x)$, $x\in\Lambda$, and $\lambda(s,x)$, $x\in \partial\Lambda$, $s\in[0,1]$.
The transformation is then realized by
\begin{equation}\label{slow}
\left\{
\begin{array}{l}
E^\tau(t)=E\left(t /\tau \right), \\
\lambda^\tau(t)=\lambda\left( t/\tau \right),
\end{array}
\right. t\in [0,\tau].
\end{equation}
The asymptotics in which we are interested is for $\tau$ large
compared to the typical relaxation time of the system, corresponding to slow transformations.
Let $\rho^\tau(t)$ and $j^\tau(t)$, $0\leq t\leq \tau$, be the
solution to the hydrodynamic equations \eqref{2.1}, \eqref{2.2} and \eqref{2.3}
with the slow external field $E^\tau$ and chemical potential $\lambda^\tau$,
that is
\begin{equation}
\left\{
\begin{array}{l}
\partial_t \rho^\tau + \nabla \cdot J(t/\tau,\rho^\tau(t))=0,\\
j^\tau(t)=J(t/\tau,\rho^\tau(t))\\
f'(\rho^\tau(t))\big|_{\partial \Lambda}=\lambda^\tau(t)
\end{array}
\right.
\label{eq-delta}
\end{equation}
where we recall that
$J(t,\rho)=-D(\rho)\nabla \rho+\chi(\rho)E(t)$.

For $s\in[0,1]$, let $\bar \rho(s)$
be the unique stationary solution of the hydrodynamics
with external field $E(s)$ and chemical potential $\lambda(s)$.
When $\tau$ is large the solution $(\rho^\tau,j^\tau)$ has an
expansion of the type
\begin{equation}\label{davide}
\begin{array}{l}
\displaystyle{
\rho^\tau (\tau s)=\bar \rho(s)+\tfrac 1\tau \, r(s)
+ o\big(\tfrac1{\tau}\big)\,,
} \\
\displaystyle{
j^\tau(\tau s)
=J(s,\bar\rho(s))+\tfrac 1\tau \, g(s) + o\big(\tfrac 1{\tau} \big)
\,.}
\end{array}
\end{equation}
By \eqref{eq-delta} we get the corresponding linear evolution
equations for the first order correction $(r,g)$,
\begin{equation}\label{davide2bis}
\left\{
\begin{array}{l}
\partial_s  \bar \rho(s) + \nabla \cdot g(s) =0 \\
g(s)= -
\nabla\cdot\big( D(\bar\rho(s)) r(s) \big)
+ r(s) \chi'(\bar \rho(s)) E(s)
\\
r(s, x)=0, \; x\in\partial \Lambda
\end{array}
\right.
\end{equation}
where we use the notation that, for a matrix $A=(a_{ij}(x))_{i,j=1}^n$,
$\nabla\cdot A$ is the vector with $i$-th coordinate $\sum_j\partial_{x_j}a_{ij}(x)$.
Note that the system \eqref{davide2bis}
has the form of a Poisson equation for $r(s)$.

Recalling the definition \eqref{W=} of the work, by evaluating the energy balance \eqref{04} along the
transformation $(\rho^\tau, j^\tau)$, we obtain
\begin{equation}
\label{lunga}
  \begin{split}
& F\big(\rho^\tau (\tau)\big)-F\big(\rho^\tau(0)\big)
=
\tau \int_0^1\!ds \int_\Lambda \!dx\,
j^\tau(\tau s)
\cdot E(s)
\\
& -\tau \int_0^1\!ds
\int_{\partial \Lambda} \! d\sigma \, \lambda(s)
j^\tau (\tau s)
\cdot \hat n
 \; \; -\tau
 \int_0^1\!ds \int_\Lambda \!dx \,
j^\tau (\tau s)
\cdot
\chi\big(\rho^\tau(\tau s)\big)^{-1}
j^\tau (\tau s).
  \end{split}
\end{equation}
In this equation $\rho^\tau(0)$ and $\rho^\tau(\tau)$ are the initial
condition for the hydrodynamic equation \eqref{eq-delta} with the
external drivings given in \eqref{slow} and the corresponding value of
the density at time $\tau$. Due to the finite relaxation time of the
system, $\rho^\tau(0)$ and $\rho^\tau(\tau)$ are not the stationary
density profiles associated to the drivings $(\lambda(0),E(0))$ and
$(\lambda(1), E(1))$. At order $1/\tau$, the difference between
$\rho^\tau(0)$ and $\bar\rho(0)$ is obtained by solving the equation
\eqref{davide2bis} for $s=0$ (here $\partial_s \bar\rho(s)$ plays the
role of a given source). The analogous statement holds for the
difference between $\rho^\tau(\tau)$ and $\bar\rho(1)$. Observe that
in this formulation the value of the density at time $0$ and $\tau$
can be exchanged so that a slow transformation from the final to the
initial state can be obtained by time reversal of the protocol.

We can analyze the equation \eqref{lunga} at the different orders in $1/\tau$,
obtaining an identity for each order.
Direct computations yield that 
at order $\tau$ the right hand side of \eqref{lunga} vanishes.

At order $\tau^0$ we get the first non trivial relationship,
\begin{equation}
\label{1t}
  \begin{split}
F\big(\bar \rho(1)\big)-F\big(\bar \rho(0)\big)
    &\; = \int_0^1\!ds \int_\Lambda\!dx \, E(s)\cdot g(s) -\int_0^1\!ds
    \int_{\partial \Lambda}\! d\sigma \,\lambda(s) g(s)\cdot \hat n
    \\
    &\;\; + \int_0^1\!ds \int_\Lambda \!dx\, r(s)
    J(s,\bar\rho(s))\cdot (\chi^{-1})'\big(\bar \rho(s)\big)
    J(s,\bar\rho(s))
    \,,
  \end{split}
\end{equation}
where we used that $\rho^\tau(0) =\bar\rho(0) + O(1/\tau)$ and
$\rho^\tau(\tau) =\bar\rho(1) + O(1/\tau)$.  
The above relation connects the variation of the free energy to the first order corrections
to the solutions of the hydrodynamic equations.
We observe that, if we consider transformations between two equilibrium states,
the last term in the right hand side of \eqref{1t}
vanishes when the intermediate states are also of equilibrium so that
$J(s,\bar\rho(s))=0$.  However the transformation can go
through nonequilibrium intermediate states.

\subsection*{Quantitative analysis of Clausius inequality}

Consider the equation \eqref{Weff2} which expresses the energy balance in the time
interval $[0,\tau]$.
Recall that the last term vanishes in the quasi-static limit.
We now compute its asymptotics when $\tau$ is large and for a slow transformation given, 
as in \eqref{slow}, in terms of a protocol $(\lambda(s),E(s))$, $s\in[0,1]$.

Rewrite equation \eqref{Weff2} for a slow transformation,
\begin{equation}\label{skype1}
\begin{split}
& 
W^{\mathrm{ren}}_{[0,\tau]}-
\big[
F(\rho^\tau(\tau))-F(\rho^\tau(0))
\big]
\\
& =
\int_0^\tau\!dt\int_\Lambda \! dx \, 
J_S(t/\tau,\rho^\tau(t))\cdot\chi(\rho^\tau(t))^{-1}J_S(t/\tau,\rho^\tau(t)).
\end{split}
\end{equation}
Recalling \eqref{Ons-form}, the symmetric part of the current is
\begin{equation}\label{symm}
J_S(s,\rho)
=-\chi(\rho)\nabla \frac{\delta V_{\lambda(s),E(s)} (\rho)}{\delta \rho}
\end{equation}
where $V_{\lambda(s),E(s)}$ is the quasi-potential associated to
$(\lambda(s), E(s))$ (we regard $s$ here as a fixed parameter).
In view of \eqref{davide}, the symmetric current has the expansion
\begin{equation}\label{symm2}
J_S(s,\rho^\tau(\tau s) )
=-\frac 1 \tau
\chi(\bar\rho(s))\nabla\left(C^{-1}_s r(s)\right)
+ O\big(\tfrac 1{\tau^2} \big).
\end{equation}
where $C_s^{-1}$ is the linear operator with integral kernel
\begin{equation}\label{C}
C^{-1}_s(x,y)=
\frac{\delta^2V_{\lambda(s),E(s)}(\bar\rho(s))}{\delta\rho(x)\delta \rho(y)}.
\end{equation}
Hence, 
\begin{equation}
\label{1taub}
W^{\mathrm{ren}}_{[0,\tau]}-
\big[ F(\rho^\tau(\tau))-F(\rho^\tau(0))\big]
=
\frac 1\tau \,B   + O \big(\tfrac 1{\tau^2}\big)
\,,
\end{equation}
where the \emph{excess} functional $B$ is:
\begin{equation}
\label{1tau}
B = \int_0^1\!ds\!\int_\Lambda\!dx \, 
\nabla\big(  C_s^{-1} r(s) \big) \cdot 
\chi(\bar\rho(s))   \nabla\big(  C_s^{-1} r(s) \big).
\end{equation}
For a transformation between and through equilibrium states, $W^{\mathrm{ren}}_{[0,\tau]}$
coincides with the total work $W_{[0,\tau]}$.
Hence, the inequality $B\ge 0$ is a restatement of the second
principle of thermodynamics
and \eqref{1taub} expresses a quantitative version of the Clausius inequality.
Note that, in the limit $\tau\to \infty$, all protocols
realize the equality $ W= \Delta F$. 
On the other hand, for finite time $\tau$,  this identity cannot be achieved
and we can select an optimal protocol by minimizing $B$. 

\section{Optimal transformations between equilibrium states}

We consider, for simplicity,
a system in one space dimension, in the domain $\Lambda=[-1,1]$,
with diffusion coefficient $D(\rho)$ and mobility $\chi(\rho)$.
Since $\bar \rho(s)$ is determined by $(\lambda(s), E(s))$ through \eqref{05},
the excess \eqref{1tau} is a functional $B=B(\lambda,E)$
of the protocol $(\lambda(s),E(s))$, $s\in[0,1]$.
In \eqref{1tau} $r(s)=r_{\lambda(s),E(s)}(x)$ is obtained by solving the following Poisson equation,
derived from \eqref{davide2bis}:
\begin{equation}\label{davide2tris}
\left\{
\begin{array}{l}
\partial_s  \bar \rho(s)
=
\Delta 
\big(
D(\bar\rho(s))  r(s) 
\big)
- \nabla\big(
\chi'(\bar \rho(s)) E(s) r(s)
\big)
\\
r(s,\pm1)=0
\,.
\end{array}
\right.
\end{equation}
Given an initial state $(\lambda_0,E_0)$ and a final state $(\lambda_1,E_1)$,
we want to minimize the excess $B(\lambda,E)$ in \eqref{1tau} as a functional of the protocol,
with the constraints $(\lambda(0),E(0))=(\lambda_0,E_0)$
and $(\lambda(1),E(1))=(\lambda_1,E_1)$.

This problem is already relevant when the initial and final states 
$(\lambda_0,E_0)$ and $(\lambda_1,E_1)$ are equilibrium states.
It appears reasonable that, in this case, 
an optimal protocol will pass through equilibrium states
$(\lambda(s),E(s))$ at every time $s$.
We will show that this is indeed the case.
Moreover, an optimal protocol can be obtained as follows.
Solve the system of partial differential equations 
\begin{equation}\label{optimal}
\left\{\begin{array}{l}
\vphantom{\Big)}
\partial_s\bar\rho(s)+\frac12\nabla(\chi(\bar\rho(s))\nabla\bar\pi(s)) = 0 \\
\vphantom{\Big)}
\partial_s\bar\pi(s) + \frac14\chi^\prime(\bar\rho(s))(\nabla\bar\pi(s))^2 = 0\\
\vphantom{\Big)}
\bar\rho(0)=\bar\rho_{\lambda_0,E_0}
\,,\,\,
\bar\rho(1)=\bar\rho_{\lambda_1,E_1}
\,,\,\,
\bar\pi(s,\pm1)=0
\,,
\end{array}\right.
\end{equation}
in the unknown $\bar\rho(s)=\bar\rho(s,x),\bar\pi(s)=\bar\pi(s,x)$,
$(s,x)\in[0,1]\times\Lambda$.
Set 
\begin{equation}
  \label{optr}
E(s)=\frac{D(\bar\rho(s))}{\chi(\bar\rho(s))}\nabla\bar\rho(s)
\,\,,\,\,\,\,
\lambda(s,\pm1)=f^\prime(\bar\rho(s,\pm1))
\,,
\end{equation}
which corresponds to the choice $G=0$ in \eqref{per-alberto}. Equation \eqref{optr} defines a
transformation between and through equilibrium states, the
corresponding minimal value of the excess functional is then given by  
\begin{equation}\label{cost-min}
B_{\mathrm{opt}} = \frac14 \int_0^1 ds\int_\Lambda dx\, \chi(\bar\rho(s)) 
\big(\nabla\pi(s)
\big)^2
\,.
\end{equation}

We emphasize that both the stationary equations \eqref{optimal} and
the corresponding minimal excess $B_{\mathrm{opt}}$ in
\eqref{cost-min} do not depend on the diffusion coefficient $D$.
In fact, in the rescaled time $s=\frac t\tau$, and in the asymptotics
$\tau\to\infty$, the system relaxes instantaneously, and therefore the
value of $D$ becomes irrelevant.

\subsubsection*{Remark on boundary conditions}

Note that in the minimization of $B$ we have not fixed the value of
$r(s)$ at $s=0$ or $1$, which corresponds, in terms of the unscaled
time variable $t$, to fix the values of the initial and final density
profiles $\rho^\tau(0)$ and $\rho^\tau(\tau)$ only at the order $1$,
and not at the order $\frac1\tau$.  On the other hand, we claim that
optimizing $B$ with the added constraints $r(0)=r(1)=0$ the infimum
does not change.  Indeed, we can consider a sequence of protocols that
are constant in time in the time intervals $[0,\epsilon]$ and
$[1-\epsilon,1]$, and close to the optimal protocol for
$s\in[\epsilon,1-\epsilon]$.  In the limit $\epsilon\to0$ the
corresponding value of $B$ approaches $B_{\mathrm{opt}}$.  As a
consequence, we deduce that, for all protocols $(\lambda(s),E(s))$,
\begin{equation}
  \label{eq:3.11.new}
W^{\mathrm{ren}}_{[0,\tau]}-(F(\bar\rho_1)-F(\bar\rho_0))\geq\frac1\tau B_{\mathrm{opt}}
+  O \big( \tfrac{1}{\tau^2} \big) \,,
\end{equation}
provided that $\rho^\tau(0)=\bar\rho_0$ and
$\rho^\tau(\tau)=\bar\rho_1$.  The previous equation differs from
\eqref{1taub} since there the variation of the free energy is 
$F(\rho^\tau(\tau))-F(\rho^\tau(0)) $.
Equality in \eqref{eq:3.11.new} can be achieved by the
limiting procedure described above.

\subsection*{Change of variables}

It will be convenient to perform a change of variables in the space of states.
Given a state $(\lambda,E)$, we associate to it 
the pair density-current $(\bar\rho,\bar J)$,
where $\bar\rho=\bar\rho_{\lambda,E}$ is the stationary density profile
defined by equation \eqref{05},
and $\bar J=-D(\bar\rho)\nabla\bar\rho+\chi(\bar\rho)E$ is the corresponding stationary current.
The correspondence $(\lambda,E)\mapsto(\bar\rho,\bar J)$ is one-to-one
and the inverse map $(\bar\rho,\bar J)\mapsto(\lambda,E)$ is given by
\begin{equation}\label{change}
\lambda(\pm1)=f^\prime(\bar\rho(\pm1))
\,,\qquad
E=\frac1{\chi(\bar\rho)}\big(D(\bar\rho)\nabla\bar\rho+\bar J\big)
\,.
\end{equation}
Observe that, since we are in one space dimension,
$\bar J$ is constant in $x$.
Under this change of variables,
equilibrium states $(\lambda,E)$ 
correspond to elements $(\bar\rho,0)$
with vanishing current.

In the new variables, the quasi-potential $V=V(\bar\rho,\bar J;\rho)$
becomes a functional on the set of density profiles $\rho:\,\Lambda\to\mb R_+$,
depending parametrically on $(\bar\rho,\bar J)$.
For $\bar J=0$ it is the local functional \eqref{localV},
\begin{equation}\label{localVb}
V(\bar\rho,0;\rho)
=
\int_\Lambda dx
\big(
f(\rho)-f(\bar\rho)-f^\prime(\bar\rho)(\rho-\bar\rho)
\big)
\,.
\end{equation}
While for arbitrary current $\bar J\in\mb R$,
the quasi-potential solves the Hamilton-Jacobi equation \eqref{r15-2},
that in the present variables reads
\begin{equation}\label{HJb}
\int_\Lambda dx\,
\chi(\rho)
\Big(\nabla\frac{\delta V(\bar\rho,\bar J;\rho)}{\delta\rho}\Big)
\Big(
\nabla\Big(
\frac{\delta}{\delta\rho}\big(V(\bar\rho,\bar J;\rho)-V(\bar\rho,0;\rho)\big)
\Big)
+\frac{\bar J}{\chi(\bar\rho)}
\Big)=0\,,
\end{equation}
where we used the Einstein relation \eqref{r29}.

In the present variables, the excess functional \eqref{1tau} becomes
\begin{equation}\label{costb}
B = \int_0^1 ds\int_\Lambda dx\, \chi(\bar\rho(s,x)) 
\Big(\nabla_x
\int_\Lambda dy
\frac{\delta^2 V (\bar\rho(s),\bar J(s);\bar\rho(s))}{\delta\rho(x)\delta\rho(y)} 
\, r(s,y)
\Big)^2
\,,
\end{equation}
where $r(s)=r(\bar\rho(s),\partial_s\bar\rho(s),\bar J(s);x)$ solves
\begin{equation}\label{davide2bisb}
\left\{
\begin{array}{l}
\partial_s  \bar \rho(s)
=
\nabla\Big(
\chi(\bar \rho(s))
\nabla\Big(
\frac{D(\bar\rho(s))}{\chi(\bar\rho(s))}r(s)
\Big)
-
\frac{\chi^\prime(\bar\rho(s))}{\chi(\bar\rho(s))}r(s)\bar J(s)
\Big)
\\
r(s,\pm1)=0
\,.
\end{array}
\right.
\end{equation}

If the initial and final states are in equilibrium, then an optimal
protocol consists of a family of equilibrium states
$(\bar\rho(s),0),s\in[0,1]$.  This will be shown by proving that
the excess functional $B$ in \eqref{costb} satisfies
\begin{equation}\label{claim1}
\frac{\delta B(\bar\rho,\bar J)}{\delta\bar J(s)}\Big|_{\bar J=0}=0
\,,\quad s\in[0,1]
\,.
\end{equation}
Indeed, this condition guarantees that stationary paths
$(\bar\rho(s),\bar J(s))$, $s\in[0,1]$, of the excess functional
$B(\bar\rho,\bar J)$ can be obtained as $(\bar\rho(s),0)$,
$s\in[0,1]$, where $\bar\rho(s)$ is a stationary path for the
functional $B(\bar\rho,0)$.
The proof of \eqref{claim1} is detailed in Appendix \ref{s:app}.

\subsection*{Hamiltonian structure}

For transformations between equilibrium states, 
in view of \eqref{claim1}, we can restrict the functional $B$ to transformations
through equilibrium states $(\bar\rho(s),0)$, $s\in[0,1]$.
Under this assumption, the excess functional $B$ \eqref{costb}
can be rewritten as
\begin{equation}\label{costc}
B = \int_0^1 ds\int_\Lambda dx\, \chi(\bar\rho(s)) 
\Big(\nabla
\Big(
\frac{D(\bar\rho(s))}{\chi(\rho(s))}
r_0(\bar\rho(s),\dot{\bar\rho}(s))
\Big)
\Big)^2
\,.
\end{equation}
By introducing 
\begin{equation}\label{alpha}
\pi(s,x)=\pi(\bar\rho(s),\dot{\bar\rho}(s);x)
=
-2\frac{D(\bar\rho(s,x))}{\chi(\bar\rho(s,x))}
r_0(\bar\rho(s),\dot{\bar\rho}(s);x)
\,,
\end{equation}
the excess functional $B$ can be written as
\begin{equation}\label{costd}
B = \frac14 \int_0^1 ds\int_\Lambda dx\, \chi(\bar\rho(s)) 
\big(\nabla\pi(s)
\big)^2
\,.
\end{equation}
and equation \eqref{davide2bisb} translates to the 
following equation for $\pi$:
\begin{equation}\label{davide2bisd}
\partial_s \bar \rho(s)
+\frac12
\nabla\Big(
\chi(\bar \rho(s))
\nabla
\pi(s)
\Big)
=
0
\,\,,\,\,\,\,
\pi(s,\pm1)=0
\,.
\end{equation}

In the form \eqref{costd}, the excess functional $B$ can be
interpreted as the action functional associated to the Lagrangian
\begin{equation}\label{lagr}
\mc L(\bar\rho,\dot{\bar\rho})
= 
\frac14\int_\Lambda dx\, \chi(\bar\rho(x)) 
\big(\nabla_x
\pi
(\bar\rho,\dot{\bar\rho};x)
\big)^2
\,.
\end{equation}
The corresponding Hamiltonian is
\begin{equation}\label{hamilt}
\mc H(\bar\rho,\bar{\pi})
= 
\sup_{\dot{\bar\rho}}
\Big\{
\int_\Lambda\!dx\,
\bar\pi\dot{\bar\rho}-\mc L(\bar\rho,\dot{\bar\rho})
\Big\}
=
\frac14\int_\Lambda dx\, \chi(\bar\rho) 
\big(\nabla_x
\bar\pi
\big)^2
\,.
\end{equation}
A straightforward computation shows that \eqref{optimal} are the Hamiltonian equations
for \eqref{hamilt}.
Note that, apart for a factor $\frac14$, \eqref{hamilt} coincides with the
Hamiltonian of the macroscopic fluctuation theory \cite[Sec.IVB]{rmp}
in the degenerate case $D=0$ and $E=0$.

\section{Explicit minimizers}

\subsection*{Optimal transformations through homogeneous equilibria}

We start by discussing how the excess functional $B$ can be minimized if
we restrict it to transformations through homogeneous equilibrium
states.  Namely, we consider $B$ in
\eqref{costd}-\eqref{davide2bisd} as a functional on paths
$\bar\rho(s)$, $s\in[0,1]$, constant in $x$ (which corresponds to
having zero external field: $E(s)=0$).

Within this setting equation \eqref{davide2bisd} for $\pi(s,x)=\pi(\bar\rho(s),\dot{\bar\rho}(s);x)$ becomes
\begin{equation}\label{homog1}
\Delta
 \pi(s,x)
=
-2\frac{\dot{\bar \rho}(s)}{\chi(\bar \rho(s))}
\,\,,\,\,\,\,
\pi(s,\pm1)=0
\,,
\end{equation}
whose solution is
\begin{equation}\label{homog2}
\pi(s,x)
=
\frac{\dot{\bar \rho}(s)}{\chi(\bar \rho(s))}(1-x^2)
\,.
\end{equation}
In view of \eqref{homog2},
the functional $B$ \eqref{costd}, restricted to homogeneous density protocols $\bar\rho(s)$ $s\in[0,1]$,
becomes
\begin{equation}\label{costhomog}
B 
= 
\frac23
\int_0^1 ds 
\frac{(\dot{\bar \rho}(s))^2}{\chi(\bar \rho(s))}
\,.
\end{equation}
Letting
$$
\Phi(\rho)=\int^\rho\!d\alpha\,\frac1{\sqrt{\chi(\alpha)}}
\,,
$$
we have
\begin{equation}\label{costhomog2}
B 
= 
\frac23
\int_0^1\! ds\, 
\big(\partial_s\Phi(\bar\rho(s))\big)^2
\,.
\end{equation}
Hence, the minimizer of this functional is obtained when
\begin{equation}\label{brachist}
\partial_s\Phi(\bar\rho(s))
=
\frac{\dot{\bar\rho}(s)}{\sqrt{\chi(\bar\rho(s))}}
=
\Phi(\bar\rho_1)-\Phi(\bar\rho_0)
\,.
\end{equation}
Thus the minimal excess
(minimizing among the homogeneous protocols) is:
\begin{equation}\label{min-homog-cost}
B_{\mathrm{opt}}=\frac23 \Big[\Phi(\bar\rho_1)-\Phi(\bar\rho_0)\Big]^2
\,.
\end{equation}
The protocol \eqref{brachist}
corresponds to the one obtained in \cite[Eq.(18)]{sc}
in the context of Markov processes with finitely many degrees of freedom.
However,
the spatial structure of our setting allows to find better protocols.
In other words,
the protocol \eqref{brachist} is not a minimizer of the excess functional
\eqref{costd} without the constraint of transformations through homogeneous equilibrium states.
Indeed, the function $\pi(t,x)$ in \eqref{homog2}
does not solve Hamiltonian equation \eqref{optimal}.
In fact we get
$$
\partial_s\pi(s,x)+\frac14\chi^\prime(\bar\rho(s))(\nabla\pi(s,x))^2
=
\frac12\chi^\prime(\bar\rho(s))\Big(\frac{\dot{\bar\rho}(s)}{\chi(\bar\rho(s))}\Big)^2(3x^2-1)
\,,
$$
which does not vanish unless $\chi$ is constant
(as in the so-called Ginzburg-Landau model).
This means, in particular, that the optimal protocol
will not be a sequence of homogeneous equilibrium states.
In the case of ideal gases the actual minimizer will be found next.

\subsection*{Ideal gas}

In the case $\chi(\rho)=\rho$, e.g., for ideal gases,
the Hamilton equation \eqref{optimal} reads
\begin{equation}\label{optimal-ideal}
\left\{\begin{array}{l}
\vphantom{\Big)}
\partial_s\bar\rho(s)+\frac12\nabla(\bar\rho(s)\nabla\bar\pi(s)) = 0 \\
\vphantom{\Big)}
\partial_s\bar\pi(s) + \frac14(\nabla\bar\pi(s))^2 = 0\\
\vphantom{\Big)}
\bar\rho(0)=\bar\rho_{\lambda_0,E_0}
\,,\,\,
\bar\rho(1)=\bar\rho_{\lambda_1,E_1}
\,,\,\,
\bar\pi(s,\pm1)=0
\,.
\end{array}\right.
\end{equation}
In particular, the second equation is decoupled
and it admits solutions with separated variables.
In the case $\bar\rho(0)=0$ and $\bar\rho(1)=\bar\rho_1$,
as can be checked by direct computations,
the solution is as follows:
\begin{equation}\label{solut-ideal}
\bar\pi(s,x)=\frac1s(1-|x|)^2
\,\,,\qquad
\bar\rho(s,x)=\frac1s\theta(|x|+s-1)\bar\rho_1
\,.
\end{equation}
The corresponding minimal value of the excess functional is
\begin{equation}\label{minimal-B-ideal}
B_{\mathrm{opt}}=\frac23\bar\rho_1
\,.
\end{equation}
This should be compared to the minimal value of $B$ through homogeneous equilibria \eqref{min-homog-cost},
which in this case is $\frac83\bar\rho_1$,
giving a flat reduction of $75\%$.

The interpretation of the solution \eqref{solut-ideal} is the following.
At time $s=0^+$ inject the required total mass $2\bar\rho_1$ at the endpoints of the domain,
giving a positive contribution to the functional $B$.
Then switch on the field 
$E=D(\bar\rho)\frac{\nabla\bar\rho}{\bar\rho}$,
which is concentrated at the points $x=\pm(1-s)$,
so that the density profile $\bar\rho$ remains a step function at all times.
Observe that the field $E$ is opposite to the current,
so the work done by the field is negative,
and thus it gives a negative contribution to the excess functional $B$.

\subsection*{Ginzburg-Landau}
This model has a constant mobility $\chi(\rho)=c$. In this case equations equations \eqref{optimal}
are linear and the solution is immediate
\begin{equation}
  \left\{
  \begin{array}{l}
    \bar\rho(x,s)=  \bar\rho_{\lambda_0,E_0}+s\left(\bar\rho_{\lambda_1,E_1}-\bar\rho_{\lambda_0,E_0}\right)\,, \\
     \pi(s,x)=\left(\bar\rho_{\lambda_1,E_1}-\bar\rho_{\lambda_0,E_0}\right)(x^2-1)\,.
  \end{array}
  \right.
\end{equation}
In particular, the optimal protocol is a sequence of homogeneous equilibrium states.

\section{Conclusions}
\label{s:con}

We have reviewed, in the context of driven diffusive systems, the
macroscopic fluctuation theory approach to non-equilibrium stationary
states \cite{rmp}. In particular, we discussed the notion of
renormalized work for which it is possible to prove a meaningful
version of the Clausius inequality for transformation between
non-equilibrium states.  
In the quasi-static limit this inequality becomes an equality.

The main purpose of the present paper has
been a quantitative discussion of the energy balance for real
transformations, that is transformations lasting a large but finite
time window. 
By rescaling the time variable we have obtained a
relationship between the variation of the
equilibrium free energy evaluated at the stationary density profiles
and the first order correction to the hydrodynamic equation, see Eq.\ \eqref{1t}.
We have then introduced the excess functional $B$ (see Eq. \eqref{1taub}-\eqref{1tau}) which accounts 
for the excess of the renormalized work with respect to the variation of free energy.
Finally, we have analyzed the minimization of $B$.
It is remarkable that, for transformations between homogeneous equilibrium states
the optimal protocol for $B$ is not a sequence of homogeneous equilibrium states.
This is due to the fact that in the framework of driven diffusive systems
the spacial structure plays a nontrivial role and the system has a finite relaxation time.
This result can be compared with the optimal protocol derived in \cite{sc}
in the context of Markov processes with finitely many degrees of freedom.
The optimal protocol for $B$ has been explicitly computed in the case
of an ideal gas in 1 space dimension,
and it exhibit peculiar features.
We also mention that, from a mathematical point of view,
the optimization problem of $B$ can be recast as a suitable
optimal mass transportation problem.

For real transformations between equilibrium states, the optimality
criterion based on the minimization of the functional $B$ in
\eqref{1tau} appears a natural choice.  On the other hand, for
transformations between stationary non-equilibrium states, $B$ only
accounts for the excess of the renormalized work with respect to the
variation of free energy.
Therefore, a selection criterion based on
the minimization of $B$ in this case is meaningful when the subtracted
counter-term $\int_{0}^{T}\! dt \int_\Lambda \!dx\,
J_\mathrm{A}(t,\rho(t)) \cdot \chi(\rho(t))^{-1}
J_\mathrm{A}(t,\rho(t))$ in \eqref{Weff} can be disregarded.  This
makes sense when the energy needed to maintain the stationary states
is supplied by free unlimited sources, e.g. the solar energy.

\appendix

\section{Analysis of the excess functional}
\label{s:app}

\subsection*{Preliminary computations}
When the quasi-potential is local, i.e. it is as in \eqref{localVb},
then the operator $C^{-1}$ in \eqref{C} is diagonal, and, in particular, it
is given by
\begin{equation}\label{Veq}
C^{-1}_{eq}(\bar\rho;x,y)
:=
C^{-1}(\bar\rho,0;x,y)
=
\frac{\delta^2 V}{\delta\rho(x)\delta\rho(y)}(\bar\rho,0;\bar\rho)
=
\frac{D(\bar\rho(x))}{\chi(\bar\rho(x))}\delta(x-y)\,.
\end{equation}

The quasi-potential $V=V(\bar\rho,\bar J;\rho)$ has a minimun (equal to $0$) in $\rho=\bar\rho$,
hence is has the general form:
\begin{equation}\label{approx1}
V(\bar\rho,\bar J;\rho)
=
\frac12\int_\Lambda\!dx \int_\Lambda\!dy \, C^{-1}(\bar\rho,\bar J;x,y)(\rho(x)-\bar\rho(x))(\rho(y)-\bar\rho(y))
+o(\rho-\bar\rho)^2
\,,
\end{equation}
where $C(\bar\rho,\bar J;x,y)$ is the limiting covariance of density correlations.
By setting 
\begin{equation}\label{Gamma}
\Gamma(\bar\rho;x,y)
=
\Big(\frac{\partial}{\partial\bar J}\frac{\delta^2 }{\delta\rho(x)\delta\rho(y)}V(\bar\rho,\bar J;\rho)\Big)
\Big|_{\rho=\bar\rho,\,\bar J=0}
\,.
\end{equation}
the kernel $C^{-1}$ in \eqref{approx1} satisfies
\begin{equation}\label{approx2}
C^{-1}(\bar\rho,\bar J;x,y)
=
\frac{D(\bar\rho(x))}{\chi(\bar\rho(x))}\delta(x-y)
+\bar J \Gamma(\bar\rho;x,y)+o(\bar J)
\,.
\end{equation}
In order to prove \eqref{claim1} we need to find an equation for $\Gamma(\bar\rho;x,y)$.
From \eqref{approx1} and \eqref{approx2}, we get
\begin{equation}\label{approx3}
\frac{\delta V(\bar\rho,\bar J;\rho)}{\delta\rho(x)}
=
\frac{D(\bar\rho(x))}{\chi(\bar\rho(x))}(\rho(x)-\bar\rho(x))
+\bar J \int_\Lambda\!dy \, \Gamma(\bar\rho;x,y)(\rho(y)-\bar\rho(y))
+\cdots
\,.
\end{equation}
We use equation \eqref{approx3}
to expand the Hamilton-Jacobi equation \eqref{HJb}
at order $2$ in $(\rho-\bar\rho)$ and at order $1$ in $\bar J$.
We get, after some simple algebraic manipulation and an integration by parts:
\begin{equation}\label{HJb-approx}
\begin{array}{l}
\displaystyle{
\int_\Lambda dx\int_\Lambda\!dy \,
(\rho(x)-\bar\rho(x))
(\rho(y)-\bar\rho(y))
\frac{D(\bar\rho(x))}{\chi(\bar\rho(x))}
\nabla_x
\Big(
\chi(\bar\rho(x))
\nabla_x
\Gamma(\bar\rho;x,y)
\Big)
}\\
\displaystyle{
=
\int_\Lambda dx\,
(\rho(x)-\bar\rho(x))
\frac{\chi^\prime(\bar\rho(x))}{\chi(\bar\rho(x))}
\nabla_x
\Big(
\frac{D(\bar\rho(x))}{\chi(\bar\rho(x))}(\rho(x)-\bar\rho(x))
\Big)
\,.}
\end{array}
\end{equation}
Since $\rho(x)-\bar\rho(x)$ is arbitrary, we conclude that $\Gamma(\bar\rho;x,y)$
(which is symmetric with respect to the exchange of $x$ and $y$)
satisfies
\begin{equation}\label{gammaeq}
\begin{array}{l}
\displaystyle{
\vphantom{\Big(}
\big(L_x(\bar\rho)+L_y(\bar\rho)\big)\Gamma(\bar\rho;x,y)
=
-
\big(R_x(\bar\rho)+R_y(\bar\rho)\big)\delta(x-y)
\,,} \\
\displaystyle{
\vphantom{\Big(}
\Gamma(\bar\rho;x,y)|_{\partial(\Lambda\times\Lambda)}=0
\,,}
\end{array}
\end{equation}
where $L(\bar\rho)$ and $R(\bar\rho)$ are the differential operators defined by
\begin{equation}\label{LR}
L(\bar\rho)\psi
=
\frac{D(\bar\rho)}{\chi(\bar\rho)}
\nabla\big(
\chi(\bar\rho)
\nabla\psi
\big)
\,,\quad
R(\bar\rho)\psi
=
\frac{D(\bar\rho)}{\chi(\bar\rho)}
\nabla\Big(
\frac{\chi^\prime(\bar\rho)}{\chi(\bar\rho)}
\psi
\Big)
\,.
\end{equation}
Equation \eqref{gammaeq} is the desired equation for $\Gamma(\bar\rho;x,y)$.

\medskip
Recall that, given $s\in[0,1]$, the function $r(s):\,\Lambda\to\mb R$ defined by \eqref{davide2bisb},
depending on the variable $x\in\Lambda$,
is a functional of $\bar\rho(s):\,\Lambda\to\mb R_+$,
$\partial_s\bar\rho(s):\,\Lambda\to\mb R$, and $\bar J(s)\in\mb R$.
Namely, we should denote $r=r(\bar\rho,\dot{\bar\rho},\bar J;x)$,
and it is defined by the equation:
\begin{equation}\label{davide2bisc}
\nabla\Big(
\chi(\bar \rho)
\nabla\Big(
\frac{D(\bar\rho)}{\chi(\bar\rho)}r
\Big)
\Big)
-
\bar J
\nabla\Big(
\frac{\chi^\prime(\bar\rho)}{\chi(\bar\rho)}r
\Big)
=\dot{\bar \rho}
\,,\quad
r(\pm1)=0
\,.
\end{equation}
In the following we shall denote
\begin{equation}\label{r0}
r_0(\bar\rho,\dot{\bar\rho};x)
:=
r(\bar\rho,\dot{\bar\rho},0;x)
\end{equation}
We also introduce the new function
\begin{equation}\label{sigma}
\gamma(\bar\rho,\dot{\bar\rho};x)
:=
\frac{\partial r(\bar\rho,\dot{\bar\rho},\bar J;x)}{\partial\bar J}\Big|_{\bar J=0}
\,.
\end{equation}
In order to prove \eqref{claim1} we need to find an equation for $\gamma(\bar\rho,\dot{\bar\rho};x)$.

If we take the derivative of both sides of \eqref{davide2bisc}
with respect to $\bar J$ and we let $\bar J=0$,
we get, after multiplying both sides by $\frac{D(\bar\rho)}{\chi(\bar\rho)}$,
\begin{equation}\label{sigmaeq}
L(\bar\rho)
\Big(
\frac{D(\bar\rho)}{\chi(\bar\rho)}
\gamma(\bar\rho,\dot{\bar\rho})
\Big)
=
R(\bar\rho)
r_0(\bar\rho,\dot{\bar\rho})
\,,\,\,
\gamma(\pm1)=0
\,,
\end{equation}
where $L(\bar\rho)$ and $R(\bar\rho)$
are the differential operators \eqref{LR}.
Equation \eqref{sigmaeq} is the desired equation for $\gamma$.

\subsection*{Proof of \eqref{claim1}}

By the definition \eqref{costb} of the excess functional $B$
and equations \eqref{Veq}, \eqref{Gamma}, \eqref{r0} and \eqref{sigma},
we have
\begin{equation}\label{proof1a}
\begin{array}{l}
\displaystyle{
\frac{\delta B(\bar\rho,\bar J)}{\delta\bar J(s)}\Big|_{\bar J=0}
=
2\int_\Lambda dx\, \chi(\bar\rho(s,x)) 
\Big(
\nabla_x
\Big(
\frac{D(\bar\rho(s,x))}{\chi(\bar\rho(s,x))}
r_0(\bar\rho(s),\dot{\bar\rho}(s);x)
\Big)
\Big)
} \\
\displaystyle{
\times\Big(
\int_\Lambda dy
\nabla_x\Gamma(\bar\rho(s);x,y)
r_0(\bar\rho(s),\dot{\bar\rho}(s);y)
+
\nabla_x
\Big(
\frac{D(\bar\rho(s,x))}{\chi(\rho(s,x))}
\gamma(\bar\rho(s),\dot{\bar\rho}(s);x)
\Big)
\Big)
\,.}
\end{array}
\end{equation}
Integrating by parts and recalling the definition \eqref{LR} of the differential operator $L(\bar\rho)$,
we can rewrite the right hand side of \eqref{proof1a}, divided by $-2$, as
\begin{equation}\label{proof1b}
\begin{array}{l}
\displaystyle{
\int_\Lambda dx\int_\Lambda dy\,
r_0(\bar\rho(s),\dot{\bar\rho}(s);x)
r_0(\bar\rho(s),\dot{\bar\rho}(s);y)
L_x(\bar\rho)
\Gamma(\bar\rho;x,y)
} \\
\displaystyle{
+
\int_\Lambda dx\,
r_0(\bar\rho(s),\dot{\bar\rho}(s);x)
L_x(\bar\rho)
\Big(
\frac{D(\bar\rho(s,x))}{\chi(\rho(s,x))}
\gamma(\bar\rho(s),\dot{\bar\rho}(s);x)
\Big)
\,.}
\end{array}
\end{equation}
Using equation \eqref{gammaeq}, we get
\begin{equation}\label{proof1c}
\begin{array}{l}
\displaystyle{
\int_\Lambda dx\int_\Lambda dy\,
r_0(\bar\rho(s),\dot{\bar\rho}(s);x)
r_0(\bar\rho(s),\dot{\bar\rho}(s);y)
L_x(\bar\rho)
\Gamma(\bar\rho;x,y)
} \\
\displaystyle{
=
-
\int_\Lambda dx\,
r_0(\bar\rho(s),\dot{\bar\rho}(s);x)
R_x(\bar\rho)r_0(\bar\rho(s),\dot{\bar\rho}(s);x)
\,.}
\end{array}
\end{equation}
Moreover, using equation \eqref{sigmaeq}, we get
\begin{equation}\label{proof1d}
\begin{array}{l}
\displaystyle{
\int_\Lambda dx\,
r_0(\bar\rho(s),\dot{\bar\rho}(s);x)
L_x(\bar\rho)
\Big(
\frac{D(\bar\rho(s,x))}{\chi(\rho(s,x))}
\gamma(\bar\rho(s),\dot{\bar\rho}(s);x)
\Big)
} \\
\displaystyle{
=\int_\Lambda dx\,
r_0(\bar\rho(s),\dot{\bar\rho}(s);x)
R_x(\bar\rho)
r_0(\bar\rho(s),\dot{\bar\rho}(s);x)
\,.}
\end{array}
\end{equation}
Combining \eqref{proof1b}, \eqref{proof1c} and \eqref{proof1d},
we get \eqref{claim1}.

\medskip
\subsection*{Acknowledgments}

We thank S.\ Ruffo for useful discussions,
the Galileo Galilei Institute for Theoretical Physics for the
stimulating atmosphere during the 2014 Workshop `Advances in
nonequilibrium statistical mechanics', and INFN for financial support.


\begin{thebibliography}{99}


\bibitem{AGMMM} 
  Aurell E., Gawedzki K., Mej{\'\i}a-Monasterio C., Mohayaee R., 
  Muratore-Ginanneschi P.;
  \emph{Refined Second Law of Thermodynamics for fast random processes.}
  J. Stat Phys \textbf{147}, 487-505 (2012).

\bibitem{45} 
  Bertini L., Gabrielli D., Jona-Lasinio G.,
  Landim C.; 
  \emph{Thermodynamic transformations of nonequilibrium states.}
    J. Stat. Phys. \textbf{149}, 773-802 (2012).  

\bibitem{46}
Bertini L., D. Gabrielli, G. Jona-Lasinio, and C. Landim;
\emph{Clausius inequality and optimality of quasistatic
  transformations for nonequilibrium stationary states.}
Phys. Rev. Lett. \textbf{110}, 020601.

\bibitem{rmp} Bertini L., De Sole A., Gabrielli D., Jona-Lasinio G.,
  Landim C.;
  \emph{Macroscopic fluctuation theory,}
  Rev. Mod. Phys., \textbf{87}, 593--636 (2015).

  
\bibitem{callen}
  Callen, H.,
  \emph{Thermodynamics and an introduction to thermostatistics,
  2nd edition}, John Wiley \& sons, New York 1985.

\bibitem{hs} 
  Hatano T., Sasa S.; 
  \emph{Steady-state thermodynamics of Langevin systems.}  
  Phys. Rev. Lett. \textbf{86}, 3463 (2001).

\bibitem{kn} 
  Komatsu T., Nakagawa N.; 
  \emph{Expression for the stationary distribution in nonequilibrium
    steady states.} 
  Phys. Rev. Lett. \textbf{100}, 030601 (2008).

\bibitem{knst} 
  Komatsu T., Nakagawa N., Sasa S., Tasaki H.;
  \emph{Entropy and nonlinear nonequilibrium thermodynamic relation
    for heat conducting steady states.} 
  J. Stat. Phys. \textbf{142}, 127-153 (2011).

\bibitem{mn}
  Maes C., Neto{\v{c}}n{\'y} K., 
  \emph{A nonequilibrium extension of the Clausius heat theorem.}
  J. Stat. Phys. \textbf{154}, 188-203 (2014).

\bibitem{mj}
Mandal, D., Jarzynski C.;
\emph{Analysis of slow transitions between nonequilibrium steady states.}
Preprint  arXiv:1507.06269
 
\bibitem{GMP}
  Muratore-Ginanneschi P.; Mej\'{\i}a-Monasterio C., Peliti L., 
  \emph{Heat release by controlled continuous-time Markov jump processes.} 
  J. Stat. Phys. \textbf{150}, 181-203 (2013).

\bibitem{op} 
  Oono Y., Paniconi M.; 
  \emph{Steady state thermodynamics.}
  Dynamic organization of fluctuations (Nishinomiya, 1997).
  Progr. Theoret. Phys. Suppl. \textbf{130}, 29-44 (1998).

\bibitem{sc}
Sivak D., Crooks G.,
\emph{Thermodynamic metrics and optimal paths.}
Phys. Rev. Lett. \textbf{108}, 190602 (2012). 

\bibitem{splib}
Spohn, H., 1991, \emph{Large scale dynamics of interacting particles}
  (Springer-Verlag, Heidelberg).


\end{thebibliography}
\end{document}